# Coarse-Graining in Quantum Mechanics: Distinguishable and Indistinguishable Particles

Patrick G. Sahrmann and Gregory A. Voth*

Corresponding Author: Gregory A. Voth

Email: gavoth@uchicago.edu

*Department of Chemistry, Chicago Center for Theoretical Chemistry, James Franck Institute, and Institute for Biophysical Dynamics, The University of Chicago, Chicago, IL 60637, USA*

**ABSTRACT**

Bottom-up coarse-grained (CG) modeling expands the spatial and temporal scales of molecular simulation by seeking a reduced, thermodynamically consistent representation of an atomistic model. Developments in CG theory have largely focused on CG modeling of atomistic systems which behave classically, while CG modeling of quantum systems has remained largely unexplored. We present in this work two fundamental advances in particle-based, bottom-up CG theory for systems obeying quantum statistical mechanics. We first expand the bottom-up CG formalism to include indistinguishable quantum particles, including bosons and fermions. We next introduce a variational optimization procedure for CG model parameterization which is founded on the relative entropy minimization (REM) principle and then bridge the classical and quantum REM methods through a semiclassical expansion in terms of the Feynman path centroid. We provide numerical examples of REM CG models of distinguishable and indistinguishable quantum systems, including as examples a harmonically trapped bosonic system and liquid water. The theoretical results presented here constitute a means to accelerate simulating thermal quantum systems ranging from distinguishable particle systems at higher temperatures to quantum indistinguishable particle systems at lower temperatures.

# I. INTRODUCTION

Classical molecular dynamics (MD) enables the simulation of atomic nuclei along a potential energy surface for the calculation of thermodynamic and dynamical properties.[1-3] However, MD is limited in scope due to the computational demands of simulating larger systems and longer time scales. To alleviate this computational burden, coarse-graining (CG) methods have been developed which reduce the particle number and/or molecular resolution in representing a particular molecular system.[4-9] CG methods are generally partitioned into two design strategies: top-down and bottom-up. Top-down CG methods parametrize CG models to reproduce experimental or thermodynamic data,[10] whereas bottom-up CG methods directly parameterize CG models from sampling of the atomistic ensemble.[6, 7]

Formally, bottom-up CG methods aim to capture the effective Hamiltonian, involving the many-body potential of mean force (mbPMF), which describes the statistical behavior of the atomistic system mapped to the CG resolution.[11-15] It is generally understood that the mbPMF includes higher-order terms beyond simple pairwise interactions,[12] and hence variational methods for obtaining approximate CG force-fields have been developed, including the multi-scale coarse-graining (MS-CG)[11, 13] and relative entropy minimization (REM)[15] methods. The MS-CG method matches the forces of a trial CG potential with the atomistic forces mapped to the CG sites.[14] Assuming adequate sampling, the MS-CG method is identical to matching the forces of the CG potential with the actual forces of the mbPMF, where direct numerical calculation of the latter is computationally infeasible due to the necessity of constrained simulation. The REM method instead matches structural correlations between the CG and mapped AA models through iterative improvement of the CG force-field.[16] The MS-CG and REM methods, although differing in target functions for variational optimization, are conceptually related.[17]

While bottom-up CGing methods and models have been developed predominantly for atomistic systems which behave classically, nuclear quantum effects are often important in accurately describing molecular phenomena even at ambient temperature. Nuclear quantum effects are often crucial in explaining vibrational spectra,[18] charge transfer,[19] and biomolecular processes such as enzyme catalysis.[20] Such quantum effects including zero point energy and tunneling cannot be accounted for via classical MD simulation. This behavior can be described mathematically, e.g., in the Feynman path integral formulation of quantum statistical mechanics,[21, 22] in which the behavior of quantum systems at thermal equilibrium is isomorphic to an interacting polymer system.[23] Each quantum particle is consequently represented as a classically isomorphic collection of 'beads', 'replicas', or quasiparticles which interact via the potential energy but are also connected by a harmonic interaction term which scales with the system temperature. The degree of quantum delocalization is also contingent on the mass of the nuclei, with lighter nuclei leading to greater delocalization. A path integral Hamiltonian can be expressed from the discretized Feynman path integral, and the resulting Boltzmann distribution can subsequently be sampled via conventional MD techniques, enabling path integral molecular dynamics (PIMD) for the calculation of equilibrium averages.[24, 25]

CGing of systems which obey quantum statistical mechanics has been investigated through numerous different approaches. Particle-based CGing of quantum particles through e.g., a center-of-mass operation, in a directly analogous manner to CGing classical particles has been conducted previously and is the predominant focus of this work.[26] Additionally, the centroid molecular dynamics (CMD) method for the calculation of static and approximate real time correlation functions can be viewed as a CGing problem, in which either adiabatic separation of the centroid mode or an approximate CG force-field for the Feynman path centroid potential of mean force must be obtained.[27, 28] The CG operation in CMD maps all path integral beads representing a physical quantum particle to the centroid of those beads via a center of geometry mapping. As an alternative, one can also CG the intermediate imaginary time slices of the Feynman imaginary time path. This CG operation retains the beginning and ending points of the Feynman paths while CGing the remaining time points into a single CG 'pseudoparticle', leaving in total three CG beads per quantum particle.[29-31] For all methods described here, conventional CG methods can be used for variational optimization of the CG force-field such as the MS-CG and REM methods as well as recently developed machine-learned effective quantum force-fields.[32-35]

Under extreme (cold) conditions in temperature and pressure, the quantum statistics of particle indistinguishability becomes pertinent.[36] In this regime, the indistinguishability of bosons and fermions produces non-negligible effective 'exchange' forces which can lead to exotic macroscopic behavior such as Bose condensation due to large-scale quantum degeneracy. A paradigmatic example of the influence of exchange effects on the condensed phase is the superfluid Bose condensate properties of $He^4$ at low temperature.[37] Additional bosonic systems which exhibit pronounced quantum degeneracy include Bose-Einstein condensates and supersolids.[38, 39] Similarly, interacting fermionic systems including Wigner crystals[40, 41] and composite bosons such as Cooper pairs in superconductors[42-44] cannot be fully described without incorporating exchange effects.

Particle indistinguishability is expressed mathematically as either the symmetry or antisymmetry of the many-body wave function under particle exchange for bosons and fermions, respectively. This is represented in the Feynman path integral isomorphism through a sum over polymer topologies in which harmonic interactions are permuted between the polymers which represent the quantum particles.[23] The number of terms in this sum grows as $n!$ for an $n$-particle system; quickly becoming computationally infeasible for large-scale PIMD simulation.[45] Additionally, the fermionic path integral carries a sign change for odd-numbered permutation operations, and hence, for large-scale systems, expectation values are evaluated as the sum of large positive and negative values which are numerically difficult to converge, a notorious challenge in many-body fermionic systems, known as the fermion sign problem.[46, 47] Given these limitations, the Path Integral Monte Carlo (PIMC) method in conjunction with the Worm algorithm has long stood as the predominant method of simulating interacting bosonic systems; however this method can produce ergodicity issues in sampling extended ring polymers.[37, 48] A recurrence relation has recently been discovered for the bosonic PIMD Hamiltonian which, followed by further algorithmic refinement, has demonstrated marked gains in efficiency relative to 'naïve' bosonic PIMD simulation.[49-51] This numerical speedup enabled the first PIMD simulation of > 1000 interacting $He^4$ atoms,[50] suggesting a viable path towards simulating condensed phase systems where exchange effects are non-negligible.

While the bottom-up CG formalism has been developed for quantum distinguishable particles,[26] a formal treatment for CGing indistinguishable particles, which constitute the most fundamental representation of matter, has not been undertaken (although we do note that the CMD method has been extended to incorporate bosons and fermions[52-54]). Our primary objective in this work is therefore to establish the formal bottom-up CG theory for indistinguishable particles. Additionally, we seek to establish connections between this quantum CG formalism and the afore-mentioned recent advances in indistinguishable PIMD simulation such that both methods might be utilized synergistically. Furthermore, while the original classically MS-CG method has been recently extended to a formal quantum treatment,[26] there is no such extension to our knowledge of the formal quantum definition of the REM method for CGing. Hence, a secondary objective of this paper is to establish the quantum theory of REM and employ it for the CGing of both distinguishable and indistinguishable systems.

The remainder of this paper is organized in the following manner: In Sec. II, we introduce (A) a thermodynamic consistency condition for CG modeling of indistinguishable particle systems and (B) the derivation of a quantum REM algorithm for variational optimization of CG models, as well as derive a semiclassical expansion to bridge its quantum and classical definitions. In Sec. III, we provide numerical examples for CGing of bosonic and distinguishable molecular systems via the REM method. Lastly, in Sec. IV, we discuss possible future directions in CGing indistinguishable particle systems as well as the potential utility in the semiclassical expansion of the quantum REM method, followed by concluding remarks in Sec. V.

## II. THEORY AND METHODOLOGY

### A. Thermodynamic Consistency Condition for Coarse-Graining Indistinguishable Particles

We assume an atomistic or more generally fine-grained (FG), non-relativistic description of an $n$-particle system with a collection of masses $(m_1, m_2, \cdots, m_n)$ whose dynamics is governed by a Hamiltonian operator, $\hat{h}$, given by

$$\hat{h} = \sum_{i=1}^{n} \frac{\hat{\mathbf{p}}_i^2}{2m_i} + \hat{u}(\hat{\mathbf{r}}) \tag{1}$$

where $\hat{\mathbf{r}} = (\hat{\mathbf{r}}_1, \hat{\mathbf{r}}_2, \cdots \hat{\mathbf{r}}_n)$ and $\hat{\mathbf{p}} = (\hat{\mathbf{p}}_1, \hat{\mathbf{p}}_2, \cdots \hat{\mathbf{p}}_n)$ are the position and momentum operators and $\hat{u}$ is the FG potential energy operator.

We will first assume that the FG system consists of distinguishable particles. The FG system is also assumed to be in thermodynamic equilibrium in the constant NVT ensemble, such that thermal averages are computed using the normalized thermal density matrix operator

$$\hat{\rho}_D = \frac{e^{-\beta \hat{h}}}{z} \tag{2}$$

where $z = \text{Tr}[e^{-\beta \hat{h}}]$. Equilibrium expectation values for observables, $\mathcal{O}$, can be expressed as $\langle \mathcal{O} \rangle = \text{Tr}[\hat{\rho} \hat{\mathcal{O}}]$. The expectation values of observables are fully described from the following matrix elements.

$$\langle \mathcal{O} \rangle = \iint d\mathbf{r}\, d\mathbf{r}'\, \langle \mathbf{r}|\hat{\rho}_D|\mathbf{r}'\rangle \langle \mathbf{r}'|\hat{\mathcal{O}}|\mathbf{r}\rangle \qquad (3)$$

where $\int d\mathbf{r}$ is a shorthand notation for integration over all $\mathbf{r}_n$ particles and $|\mathbf{r}\rangle = |\mathbf{r}_1 \cdots \mathbf{r}_n\rangle$.

As we are interested in CG representations, we next limit ourselves to observables which depend only on a reduced representation of the FG degrees of freedom, which we denote as the CG variables $\mathbf{R} = (\mathbf{R}_1, \mathbf{R}_2, \cdots \mathbf{R}_N)$ and $\mathbf{P} = (\mathbf{P}_1, \mathbf{P}_2, \cdots \mathbf{P}_N)$. The FG and CG configurations of a system are connected through a set of $N$ mapping functions, $\mathcal{M}$, such that $\mathcal{M}_I(\mathbf{r}) = \mathbf{R}_I$ for $I = 1 \dots N$. Similarly, mapping operators in momentum space, $\mathcal{P}$, can be defined such that $\mathcal{P}_I(\mathbf{p}) = \mathbf{P}_I$ for $I = 1 \dots N$. The position and momentum mapping functions are related.[13] The ensemble average of observables which depend on this CG description, $\mathcal{O}(\mathbf{R}, \mathbf{P})$, can be evaluated in the position basis as

$$\langle \mathcal{O} \rangle = \iint d\mathbf{R}\, d\mathbf{R}'\, p(\mathbf{R}, \mathbf{R}') \mathcal{O}(\mathbf{R}', \mathbf{R}) \qquad (4)$$

where in Eq. (4) we have defined a position-basis distribution for the CG variables

$$p(\mathbf{R}, \mathbf{R}') \equiv \iint d\mathbf{r}\, d\mathbf{r}'\, \langle \mathbf{r}|\boldsymbol{\delta}(\widehat{\mathcal{M}}(\hat{\mathbf{r}}) - \mathbf{R})\hat{\rho}_D \boldsymbol{\delta}(\widehat{\mathcal{M}}(\hat{\mathbf{r}}) - \mathbf{R}')|\mathbf{r}'\rangle \qquad (5)$$

and we have introduced notation for the matrix elements of operator $\hat{\mathcal{O}}$ such that $\hat{\mathcal{O}}_{\mathbf{r}'\mathbf{r}} = \mathcal{O}(\mathcal{M}(\mathbf{r}'), \mathcal{M}(\mathbf{r})) = \mathcal{O}(\mathbf{R}', \mathbf{R}) = \hat{\mathcal{O}}_{\mathbf{R}'\mathbf{R}}$. We emphasize that the results presented here are contingent on the functional form of the operator $\hat{\mathcal{O}}$ being equal between the FG and CG descriptions. This is identical to the statement that the observable $\mathcal{O}$ is fully described by the CG degrees of freedom. If $\mathcal{O}$ is described by FG degrees of freedom, a representability issue arises[55] wherein the equality of the operator matrix elements when evaluated in the FG and CG Hilbert spaces no longer holds.[26] The distribution in Eq. (5) is related to the reduced density matrix.[56] It is understood here that the bold Dirac delta presented in Eq. (5) is shorthand for a product of Dirac delta functions for all $N$ CG sites, i.e.,

$$\delta(\widehat{\mathcal{M}}(\hat{\mathbf{r}}) - \mathbf{R}) = \prod_{I=1}^{N} \delta(\widehat{\mathcal{M}}_I(\hat{\mathbf{r}}) - \mathbf{R}_I) \tag{6}$$

To establish thermodynamic consistency, we define a Hamiltonian which governs the exact behavior of the CG variables.

$$\widehat{H} = \sum_{I=1}^{N} \frac{\widehat{\mathbf{P}}_I^2}{2m_I} + \widehat{U}(\widehat{\mathbf{R}}) \tag{7}$$

We note that the real-time dynamics of the CG description is treated rigorously through application of the Nakajima-Zwanzig projection operator method.[57, 58] However, given that full application of this method is computationally burdensome and that the calculation of real-time dynamics of condensed phase quantum systems remains challenging due to the sign problem,[59] our intent is instead for the Hamiltonian of Eq. (7) to recapitulate all static equilibrium correlations at the CG resolution.

The CG Hamiltonian produces the following density matrix at thermal equilibrium

$$\hat{\rho}_{CG} = \frac{e^{-\beta \widehat{H}}}{Z} \tag{8}$$

where $Z = \text{Tr}[e^{-\beta \widehat{H}}]$. The expectation value of equilibrium configurational observables is then described by

$$\langle \mathcal{O} \rangle = \iint d\mathbf{R}\, d\mathbf{R}' \langle \mathbf{R} | \hat{\rho}_{CG} | \mathbf{R}' \rangle \langle \mathbf{R}' | \widehat{\mathcal{O}} | \mathbf{R} \rangle \tag{9}$$

where, again, $\int d\mathbf{R}$ is shorthand notation for all integration over all $\mathbf{R}_N$ particles and we have used the shorthand notation $|\mathbf{R}\rangle = |\mathbf{R}_1 \cdots \mathbf{R}_N\rangle$.

Thermodynamic consistency between the FG and CG descriptions necessitates the equivalence of Eqs. (4) and (9), which then implies

$$\langle \mathbf{R} | e^{-\beta \widehat{H}} | \mathbf{R}' \rangle = \iint d\mathbf{r}\, d\mathbf{r}' \langle \mathbf{r} | \delta(\widehat{\mathcal{M}}(\hat{\mathbf{r}}) - \mathbf{R}) e^{-\beta \hat{h}} \delta(\widehat{\mathcal{M}}(\hat{\mathbf{r}}) - \mathbf{R}') | \mathbf{r}' \rangle \tag{10}$$

The thermodynamic consistency established in Eq. (10) holds for distinguishable particles. To extend this to indistinguishable particles, the thermal density matrix must be (anti)symmetrized under permutation operations via the appropriate (anti)symmetrization operator, $\hat{S}$. Specifically, bosonic systems must be symmetric under particle exchange, while fermionic systems must be antisymmetric under particle exchange. We will not consider spin degrees of freedom in this work, although the theoretical results obtained here can be extended to include them. We additionally assume, reasonably, that the FG potential operator is invariant under particle exchanges, i.e., $[\hat{u}, \hat{\pi}_\sigma] = 0 \ \forall \ \sigma \in S_n$, and that this condition also holds for the CG potential operator, i.e., $[\hat{U}, \hat{\pi}_\sigma] = 0 \ \forall \ \sigma \in S_N$.

The thermal density matrix which describes $n$ indistinguishable particles, $\hat{\rho}$, is related to the thermal density matrix of the same set of $n$ distinguishable particles in the following manner $\hat{\rho} \propto \hat{S}\hat{\rho}_D$, where the proportionality constant is the ratio of the partition functions. For the indistinguishable particle case, we thus rewrite Eq. (10) as

$$\langle \mathbf{R}|e^{-\beta \hat{H}}|\mathbf{R}'\rangle = \iint d\mathbf{r}\, d\mathbf{r}'\, \langle \mathbf{r}|\delta(\widehat{\mathcal{M}}(\hat{\mathbf{r}}) - \mathbf{R})\hat{S}e^{-\beta \hat{h}}\delta(\widehat{\mathcal{M}}(\hat{\mathbf{r}}) - \mathbf{R}')|\mathbf{r}'\rangle \tag{11}$$

We now evaluate the right-hand side of Eq. (11) in the position basis and explicitly substitute for the definition of the (anti)symmetrization operator, $\hat{S}$,

$$\iint d\mathbf{r}\, d\mathbf{r}'\langle \mathbf{r}|\delta(\widehat{\mathcal{M}}(\hat{\mathbf{r}}) - \mathbf{R})\hat{S}e^{-\beta \hat{h}}\delta(\widehat{\mathcal{M}}(\hat{\mathbf{r}}) - \mathbf{R}')|\mathbf{r}'\rangle$$

$$= \frac{1}{n!}\iint d\mathbf{r}\, d\mathbf{r}' \sum_{\sigma \in S_n} \xi^{n_{\mathrm{PP}}} \langle \mathbf{r}_{\pi_\sigma(1)}, \cdots, \mathbf{r}_{\pi_\sigma(n)}|e^{-\beta \hat{h}}|\mathbf{r}'\rangle \tag{12}$$

$$\times \delta(\mathcal{M}(\mathbf{r}) - \mathbf{R})\delta(\mathcal{M}(\mathbf{r}') - \mathbf{R}')$$

where $\xi = \pm 1$ for bosons or fermions and $N_{\mathrm{PP}}$ denotes the number of pair permutations given by permutation operator $\hat{\pi}_\sigma$ for $\sigma$ belonging to the permutation group $S_n$, such that $\hat{\pi}_\sigma f(\mathbf{r}_1, \cdots, \mathbf{r}_n) = f(\mathbf{r}_{\pi_\sigma(1)}, \cdots, \mathbf{r}_{\pi_\sigma(n)})$. We next evaluate the matrix elements in Eq. (12) utilizing the Trotter product formula to establish the Feynman path integral expression in imaginary time

$$\left\langle \mathbf{r}_{\pi_\sigma(1)}, \cdots, \mathbf{r}_{\pi_\sigma(n)} \left| e^{-\beta \hat{h}} \right| \mathbf{r}' \right\rangle$$

$$= \int_{\mathbf{r}'_1(0)=\mathbf{r}'_1, \cdots, \mathbf{r}'_n(0)=\mathbf{r}'_n}^{\mathbf{r}'_1(\beta\hbar)=\mathbf{r}_{\pi_\sigma(1)}, \cdots, \mathbf{r}'_n(\beta\hbar)=\mathbf{r}_{\pi_\sigma(n)}} \mathcal{D}\mathbf{r}'(\tau) e^{-S[\mathbf{r}'(\tau)]/\hbar} \quad (13)$$

$$\times \delta(\mathcal{M}(\mathbf{r}) - \mathbf{R})\delta(\mathcal{M}(\mathbf{r}') - \mathbf{R}')$$

In Eq. (13), we have defined the FG Euclidian action

$$S[\mathbf{r}(\tau)] = \int_0^{\beta\hbar} d\tau \, \mathcal{K}[\dot{\mathbf{r}}(\tau)] + u([\mathbf{r}(\tau)]) \quad (14)$$

and the imaginary time kinetic energy

$$\mathcal{K}[\mathbf{r}(\tau)] = \frac{1}{2}\dot{\mathbf{r}}(\tau)^T \mathbf{m} \dot{\mathbf{r}}(\tau) \quad (15)$$

Note that in the case of indistinguishable particles that the mass term must be the same for all particles, i.e., $\mathbf{m} = \mathbf{I}_n m$.

We can similarly evaluate the left-hand side of Eq. (11) via the Trotter product formula to produce the following path integral for the CG matrix elements

$$\langle \mathbf{R} | e^{-\beta \hat{H}} | \mathbf{R}' \rangle = \int_{\mathbf{R}'}^{\mathbf{R}} \mathcal{D}\mathbf{R}'(\tau) \, e^{-S_{CG}[\mathbf{R}'(\tau)]/\hbar} \quad (16)$$

where we have analogously defined the CG Euclidian action

$$S_{CG}[\mathbf{R}(\tau)] = \int_0^{\beta\hbar} d\tau \, \mathcal{K}[\dot{\mathbf{R}}(\tau)] + U([\mathbf{R}(\tau)]) \quad (17)$$

and the CG imaginary time kinetic energy as

$$\mathcal{K}[\mathbf{R}(\tau)] = \frac{1}{2}\dot{\mathbf{R}}(\tau)^T \mathbf{M} \dot{\mathbf{R}}(\tau) \quad (18)$$

Substituting Eqs. (12), (13), and (16) into Eq. (11) then produces the following equivalence between the FG and CG path integrals

$$\int_{\mathbf{R}'}^{\mathbf{R}} \mathcal{D}\mathbf{R}'(\tau)\, e^{-S_{CG}[\mathbf{R}'(\tau)]/\hbar}$$

$$= \frac{1}{n!} \iint d\mathbf{r}\, d\mathbf{r}' \sum_{\sigma \in S_n} \xi^{n_{PP}} \int_{\mathbf{r}'_1(0)=\mathbf{r}'_1,\cdots,\mathbf{r}'_n(0)=\mathbf{r}'_n}^{\mathbf{r}'_1(\beta\hbar)=\mathbf{r}_{\pi_\sigma(1)},\cdots,\mathbf{r}'_n(\beta\hbar)=\mathbf{r}_{\pi_\sigma(n)}} \mathcal{D}\mathbf{r}'(\tau)\, e^{-S[\mathbf{r}'(\tau)]/\hbar} \quad (19)$$

$$\times \delta(\mathcal{M}(\mathbf{r}) - \mathbf{R})\delta(\mathcal{M}(\mathbf{r}') - \mathbf{R}')$$

The imaginary time path integral on the right-hand side places constraints on the start and endpoints for all paths through the Dirac delta functions. As in previous work,[26] we can introduce Dirac delta functions along each imaginary time slice to expand the mapping constraint to the entire imaginary time path rather than only the start and end points. This produces a double path integral of the following form

$$\int_{\mathbf{R}'}^{\mathbf{R}} \mathcal{D}\mathbf{R}'(\tau)\, e^{-S_{CG}[\mathbf{R}'(\tau)]/\hbar}$$

$$= \frac{1}{n!} \int_{\mathbf{R}'}^{\mathbf{R}} \mathcal{D}\mathbf{R}'(\tau) \iint d\mathbf{r}\, d\mathbf{r}' \sum_{\sigma \in S_n} \xi^{n_{PP}} \int_{\mathbf{r}'_1(0)=\mathbf{r}'_1,\cdots,\mathbf{r}'_n(0)=\mathbf{r}'_n}^{\mathbf{r}'_1(\beta\hbar)=\mathbf{r}_{\pi_\sigma(1)},\cdots,\mathbf{r}'_n(\beta\hbar)=\mathbf{r}_{\pi_\sigma(n)}} \mathcal{D}\mathbf{r}'(\tau)\, e^{-S[\mathbf{r}'(\tau)]/\hbar} \quad (20)$$

$$\times \delta\big(\mathcal{M}(\mathbf{r}')(\tau) - \mathbf{R}'(\tau)\big)$$

where it is understood that the single Dirac delta in Eq. (20) is shorthand notation for the product of Dirac deltas across all imaginary time slices, i.e.

$$\delta\big(\mathcal{M}(\mathbf{r}')(\tau) - \mathbf{R}'(\tau)\big) = \delta(\mathcal{M}(\mathbf{r}) - \mathbf{R})\delta(\mathcal{M}(\mathbf{r}') - \mathbf{R}') \prod_{i=2}^{P-1} \delta(\mathcal{M}(\mathbf{r}_i) - \mathbf{R}_i) \quad (21)$$

An exact expression for the CG action can then be obtained from Eq. (20)

$$S_{CG}[\mathbf{R}'(\tau)] = -\hbar$$

$$\times \ln\Bigg[\frac{1}{n!} \sum_{\sigma \in S_n} \xi^{n_{PP}} \iint d\mathbf{r}\, d\mathbf{r}' \int_{\mathbf{r}'_1(0)=\mathbf{r}'_1,\cdots,\mathbf{r}'_n(0)=\mathbf{r}'_n}^{\mathbf{r}'_1(\beta\hbar)=\mathbf{r}_{\pi_\sigma(1)},\cdots,\mathbf{r}'_n(\beta\hbar)=\mathbf{r}_{\pi_\sigma(n)}} \mathcal{D}\mathbf{r}'(\tau)\, e^{-S[\mathbf{r}'(\tau)]/\hbar}$$

$$\times \delta\big(\mathcal{M}(\mathbf{r}')(\tau) - \mathbf{R}'(\tau)\big)\Bigg] \quad (22)$$

Eq. (22) is the most general expression which can be written for the thermodynamic consistency relation of indistinguishable particle CGing, which states that the action-weighted sum of CG paths is equivalent to an action-weighted sum of (anti)symmetrized paths over the FG configurations which map to the CG configuration. Figure 1 summarizes the differences in the CG operation between classical statistical mechanics, quantum statistical mechanics, and indistinguishable quantum statistical mechanics.

Our goal is to derive an exact expression for the CG potential operator, $\hat{U}$. However, we desire that sampling of the FG description via PIMD/PIMC would enable training of the CG model, and, in theory, variational optimization of a maximally expressive CG force-field basis set would reproduce the mbPMF. Consequently, the remainder of this section will be devoted to deriving an analytical expression for the CG potential under a series of successive constraints. We first limit ourselves to closed-path evaluations, i.e., $\mathbf{r} = \mathbf{r}'$ and $\mathbf{R} = \mathbf{R}'$. This necessarily limits the applicability of the following results to strictly configuration-dependent operators. This constraint allows for the simplification of Eq. (20) in the following manner

$$\int_{\mathbf{R}}^{\mathbf{R}} \mathcal{D}\mathbf{R}'(\tau)\, e^{-S_{CG}[\mathbf{R}'(\tau)]/\hbar}$$

$$= \frac{1}{n!} \int_{\mathbf{R}}^{\mathbf{R}} \mathcal{D}\mathbf{R}'(\tau) \int d\mathbf{r} \sum_{\sigma \in S_n} \xi^{n_{\mathrm{PP}}} \int_{\mathbf{r}_1(0)=\mathbf{r}_1,\cdots,\mathbf{r}_n(0)=\mathbf{r}_n}^{\mathbf{r}_1(\beta\hbar)=\mathbf{r}_{\pi_\sigma(1)},\cdots,\mathbf{r}_n(\beta\hbar)=\mathbf{r}_{\pi_\sigma(n)}} \mathcal{D}\mathbf{r}'(\tau)\, e^{-S[\mathbf{r}'(\tau)]/\hbar} \quad (23)$$

$$\times \delta\big(\mathcal{M}(\mathbf{r}(\tau)) - \mathbf{R}(\tau)\big)$$

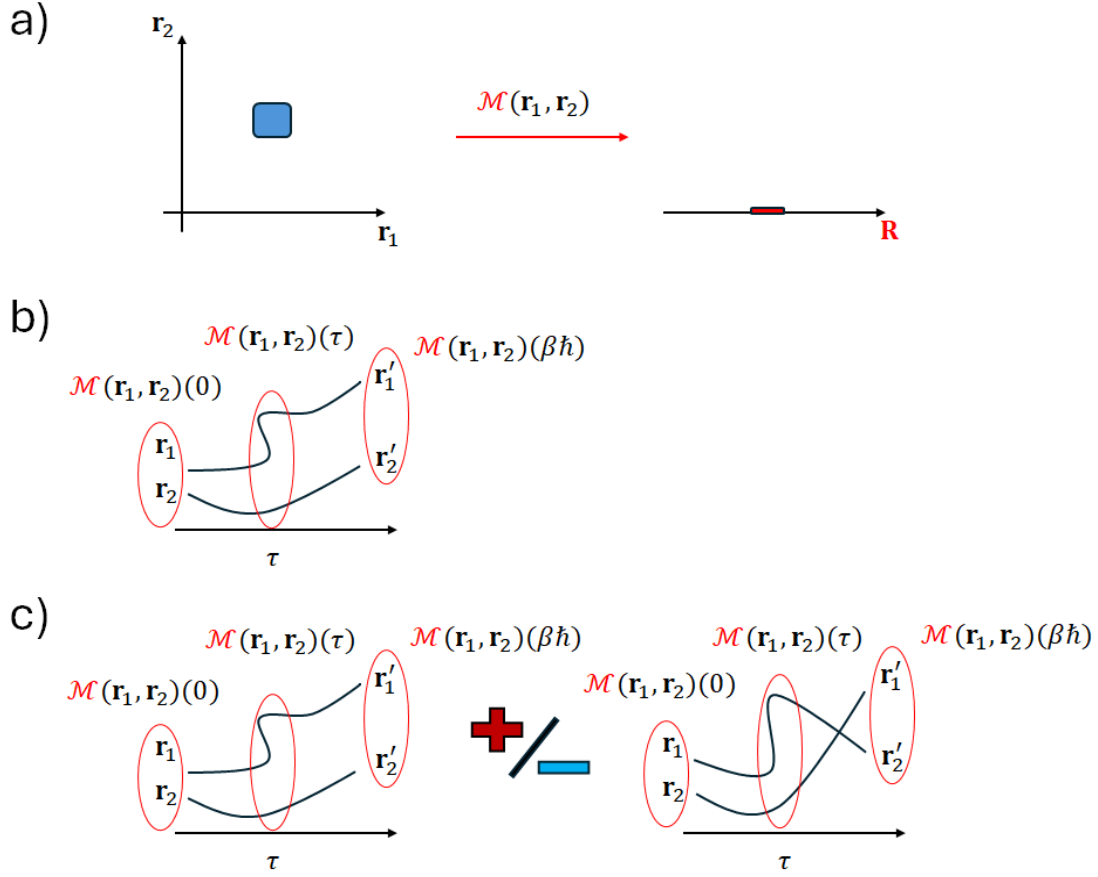

**Figure 1.** Diagram representing differences between classical and quantum CG for distinguishable and indistinguishable particles. a) Classical CG maps FG phase space regions to points on the CG phase space. b) Quantum CG of distinguishable particles maps FG trajectories in imaginary time to CG trajectories in imaginary time. c) Quantum CG of indistinguishable particles maps a sum of FG trajectories in imaginary time to CG trajectories in imaginary time, where the endpoints of each FG trajectory are permuted symmetrically or antisymetrically depending on the particle type.

We next restrict ourselves to finite time discretization for some set of time slices $P$ but recognize that the $P \to \infty$ limit should be taken to achieve a formally exact result. In the $P \to \infty$ limit, Eq. (23) can be expressed as

$$\int d\mathbf{R}^{[2\cdots P]} e^{-\beta \Phi(\mathbf{R}^{[1\cdots P]})} \propto \int d\mathbf{R}^{[2\cdots P]} \int d\mathbf{r}^{[1\cdots P]} e^{-\beta \phi(\mathbf{r}^{[1\cdots P]})} \delta^{[1\cdots P]}(\mathcal{M}(\mathbf{r}) - \mathbf{R}) \qquad (24)$$

where $\mathbf{r}_i^{[1\cdots P]}$ is shorthand notation for the $P$ replicas of the $i^{th}$ FG particle, $\mathbf{R}_I^{[1\cdots P]}$ is shorthand notation for the $P$ replicas of the $I^{th}$ CG site, and $\delta^{[1\cdots P]}$ is shorthand notation for the product of the $N \times P$ Dirac delta functions. We also have defined the FG ring polymer potential[60]

$$\phi(\mathbf{r}^{[1\cdots P]}) = \sum_{i=1}^{P}\sum_{j=1}^{n}\frac{1}{2}m\omega_P^2(\mathbf{r}_j^i - \mathbf{r}_j^{i+1})^2 + \frac{1}{P}\sum_{i=1}^{P}u(\mathbf{r}^i) - \frac{1}{\beta}\ln\mathcal{A}_n \qquad (25)$$

where $\omega_P = \sqrt{P}/\beta\hbar$ and $\mathbf{r}_j^{P+1} = \mathbf{r}_j^1 \;\forall\; j$. The term $\mathcal{A}_n$ corresponds to perm $\tilde{\mathbf{A}}_n$ or det $\tilde{\mathbf{A}}_n$ for bosonic and fermionic systems, respectively. The matrix $\tilde{\mathbf{A}}_n$ is defined such that $\tilde{A}_{n,ij} = \frac{A_{n,ij}}{A_{n,ii}}$ and the elements of matrix $\mathbf{A}_n$ are given by

$$A_{n,ij} = \exp-\frac{1}{2}\beta m\omega_P^2(\mathbf{r}_i^P - \mathbf{r}_j^1)^2 \qquad (26)$$

It is clear that $\mathbf{A}_n$ is an $n \times n$ matrix. This term is unique to indistinguishable particles and reduces to $\tilde{\mathbf{A}}_n = \mathbf{A}_n = \mathbf{I}_n$ for the case of distinguishable particles, i.e., when $\xi = 0$.

Similarly, we may express the CG ring polymer potential as

$$\Phi(\mathbf{R}^{[1\cdots P]}) = \mathcal{K}^P(\mathbf{R}^{[1\cdots P]}) + \frac{1}{P}\sum_{i=1}^{P}U(\mathbf{R}^i) \qquad (27)$$

where we have defined the discretized 'kinetic energy' term

$$\mathcal{K}^P(\mathbf{R}^{[1\cdots P]}) = \sum_{i=1}^{P}\sum_{j=1}^{N}\frac{1}{2}M\omega_P^2(\mathbf{R}_j^i - \mathbf{R}_j^{i+1})^2 \qquad (28)$$

such that $\mathbf{R}_j^{P+1} = \mathbf{R}_j^1 \;\forall\; j$. Note that if $\xi = -1$, as in the fermionic case, then $U$ may adopt complex values due to the possible negativity of the determinant term.

In principle, Eq. (24) is solvable for the CG potential. However, the suggested form of the total potential for PIMD simulation in Eq. (25) is inadequate due to the poor scalability of the permanent/determinant calculation (neglecting the sign problem involved in the latter for now). Instead, we will capitalize on the recently developed and more numerically efficient recurrence relations for bosonic and fermionic PIMD to rewrite Eq. (24). Following Refs.[49, 61], we write

$$\int d\mathbf{R}^{[2\cdots P]}\, e^{-\beta\Phi(\mathbf{R}^{[1\cdots P]})} \propto \int d\mathbf{R}^{[2\cdots P]}\int d\mathbf{r}^{[1\cdots P]}\, e^{-\beta V_n(\mathbf{r}^{[1\cdots P]})}\boldsymbol{\delta}^{[1\cdots P]}(\mathcal{M}(\mathbf{r}) - \mathbf{R}) \qquad (29)$$

In Eq. (29), we have defined a potential which 'renormalizes' the exchange effects such that

$$V_n(\mathbf{r}^{[1\cdots P]}) = -\frac{1}{\beta}\ln w_n(\mathbf{r}^{[1\cdots P]}) + \frac{1}{P}\sum_{i=1}^{P} u(\mathbf{r}^i) \tag{30}$$

and the potential term, $w_n(\mathbf{r}^n)$, can be evaluated via the following recurrence relation

$$w_n(\mathbf{r}^{[1\cdots P]}) = \frac{1}{n}\sum_{k=1}^{n} \xi^{k-1} e^{-\beta T_n^k} w_{n-k}(\mathbf{r}^{[1\cdots P]}) \tag{31}$$

where, as defined previously, $\xi = \pm 1$ for bosons or fermions, and $w_0(\mathbf{r}^{[1\cdots P]}) = 1$ by definition. In Eq. (29), $T_n^k(\mathbf{r}_{n-k+1}^{[1\cdots P]}, \ldots, \mathbf{r}_n^{[1\cdots P]})$ constitutes the spring connection terms of $n - k + 1$ to $n$ particles

$$T_n^k(\mathbf{r}_{n-k+1}^{[1\cdots P]}, \ldots, \mathbf{r}_n^{[1\cdots P]}) = \frac{1}{2}m\omega_P^2 \sum_{l=n-k+1}^{n}\sum_{i=1}^{P}(\mathbf{r}_l^{i+1} - \mathbf{r}_l^i)^2 \tag{32}$$

such that $\mathbf{r}_l^{P+1} = \mathbf{r}_{l+1}^1$ for all cases except $l = n$. In this particular case, $\mathbf{r}_n^{P+1} = \mathbf{r}_{n-k+1}^1$. From Eq. (29) we have the following expression for the total CG potential

$$\Phi(\mathbf{R}^{[1\cdots P]}) \propto -k_B T \ln\left[\int d\mathbf{r}^{[1\cdots P]} e^{-\beta V_n(\mathbf{r}^{[1\cdots P]})} \delta^{[1\cdots P]}(\mathcal{M}(\mathbf{r}) - \mathbf{R})\right] \tag{33}$$

Substituting Eq. (27) into Eq. (33), we obtain an analytical expression for the CG potential in the $P \to \infty$ limit

$$U(\mathbf{R}^{[1\cdots P]}) \propto -P \times k_B T \left(\ln\left[\frac{\int d\mathbf{r}^{[1\cdots P]} e^{-\beta V_n(\mathbf{r}^{[1\cdots P]})} \delta^{[1\cdots P]}(\mathcal{M}(\mathbf{r}) - \mathbf{R})}{e^{-\beta \mathcal{K}^P(\mathbf{R}^{[1\cdots P]})}}\right]\right) \tag{34}$$

Eq. (34) describes the potential for $N$ CG sites whose thermal density matrix will produce expectation values of position-dependent CG observables consistent with the expectation value of the same CG observable for a set of $n$ indistinguishable particles in thermal equilibrium.

Furthermore, it is clear how the CG potential changes under various approximate limiting cases. First, in the case where particles can be rendered distinguishable, one designates $\xi = 0$. This ensures all summed paths start and end at the same point in imaginary time, and the $w_n(\mathbf{r}^{[1\cdots P]})$ terms reduce to the harmonic terms employed in distinguishable PIMD. In the classical limit, $P = 1$ and thus $w_n(\mathbf{r}) = 1$ and $\mathcal{K}^1(\mathbf{R}) = 0$, returning the well-established expression for the thermodynamic consistency of a classical CG potential.[13]

**B. Quantum Coarse-Graining from a Relative Entropy Principle**

To develop an expression for the quantum analogue to the relative entropy for classical CG systems, we first define the entropy for a quantum system as the (unitless) von Neumann entropy

$$S[\hat{\rho}] \equiv - \text{Tr}[\hat{\rho} \ln \hat{\rho}] \tag{35}$$

We note that hereafter we assume that the effects of indistinguishability, when present, are already included in the definition of the density matrices. The relative entropy between the FG and CG models can then be expressed as

$$S_{\text{rel}}[\hat{\rho}||\hat{\rho}_{CG}] = H_{\mathcal{M}}[\hat{\rho}||\hat{\rho}_{CG}] - S[\hat{\rho}] \tag{36}$$

where $H_{\mathcal{M}}$ is the cross-entropy between the FG and CG density matrices. We define the cross-entropy as

$$H_{\mathcal{M}}[\hat{\rho}||\rho_{CG}] \equiv - \text{Tr}[\hat{\rho} \ln \hat{\rho}_{CG}] + \text{Tr}\left[\hat{\rho} \ln \int d\mathbf{r}' \, \delta\left(\widehat{\mathcal{M}}(\hat{\mathbf{r}}) - \mathcal{M}(\mathbf{r}')\right)\right] \tag{37}$$

The latter term in Eq. (37) denotes the volume of the CG 'macrostate' to ensure proper weighting of the CG ensemble, and can be viewed as the quantum analogue of the mapping entropy, $S_{\text{map}}$.[62-64] We therefore define this term as the mapping entropy.

$$\hat{S}_{map}(\hat{\mathbf{r}}) \equiv \ln \int d\mathbf{r}' \, \delta\left(\widehat{\mathcal{M}}(\hat{\mathbf{r}}) - \mathcal{M}(\mathbf{r}')\right) \tag{38}$$

Substituting Eqs. (37) and (38) into Eq. (36), an expression for the relative entropy as a function of the tunable CG model parameters, $\boldsymbol{\lambda}$, is then

$$S_{\text{rel}}(\boldsymbol{\lambda}) = \text{Tr}\left[\hat{\rho} \ln \hat{\rho}\right] - \text{Tr}[\hat{\rho} \ln \hat{\rho}_{CG}(\boldsymbol{\lambda})] + \text{Tr}[\hat{\rho}\hat{S}_{map}] \tag{39}$$

We note that, as in the classical relative entropy, the mapping entropy is independent of the model parameters. At the condition of thermodynamic equilibrium Eq. (39) can be evaluated using Eqs. (2) and (8) as

$$S_{\text{rel}}(\lambda) = \beta \text{Tr}\,[\hat{\rho}(\hat{H}(\lambda) - \hat{h})] - \beta(F(\lambda) - f) + \text{Tr}[\hat{\rho}\hat{S}_{map}] \tag{40}$$

where we have defined the FG and CG free energies $f$ and $F$, respectively. Eq. (40) relates the relative entropy to differences in the average and free energies of the two models; the Bogoliubov inequality is hence closely related to the relative entropy, as has been described previously.[15]

For many-body interacting systems the relative entropy cannot be calculated analytically as this entails calculation of the partition function. Instead, gradient descent methods are often employed. Note that for a model parameter $\lambda$ the following is true

$$\frac{\partial F}{\partial \lambda} = \text{Tr}\left[\hat{\rho}_{CG} \frac{\partial \hat{H}}{\partial \lambda}\right] \tag{41}$$

which can be viewed as an extension of the Hellman-Feynman theorem at finite temperature.[65] From Eq. (41) the derivative of the relative entropy with respect to a model parameter, $\lambda$, can be calculated as

$$\frac{\partial S_{\text{rel}}}{\partial \lambda} = \beta\,\text{Tr}\left[\hat{\rho}\frac{\partial \hat{H}}{\partial \lambda}\right] - \beta\,\text{Tr}\left[\hat{\rho}_{CG}\frac{\partial \hat{H}}{\partial \lambda}\right] \tag{42}$$

Similarly, the second derivative can be calculated as

$$\begin{aligned}\frac{\partial^2 S_{\text{rel}}}{\partial \lambda^2} &= \beta\,\text{Tr}\left[\hat{\rho}\frac{\partial^2 \hat{H}}{\partial \lambda^2}\right] \\ &- \beta\,\text{Tr}\left[\hat{\rho}_{CG}\frac{\partial^2 \hat{H}}{\partial \lambda^2}\right] \\ &+ \beta^2\,\text{Tr}\left[\hat{\rho}_{CG}\left(\frac{\partial \hat{H}}{\partial \lambda}\right)^2\right] - \beta^2\left(\text{Tr}\left[\hat{\rho}_{CG}\frac{\partial \hat{H}}{\partial \lambda}\right]\right)^2\end{aligned} \tag{43}$$

All terms in the first and second derivative expressions can be calculated from PIMD simulations, as employed in this work. Furthermore, when the model Hamiltonian is linear with respect to $\lambda$, then the second derivative in Eq. (43) is proportional to the variance, and therefore there exists a single global minimum for $\lambda$. Iterative gradient descent schemes can be developed from Eqs. (42) and (43) such that for the $k^{\text{th}}$ gradient step the update for model parameter $\lambda$ is

$$\lambda^{k+1} = \lambda^k - \chi \partial_\lambda S_{rel}(\lambda^k) \tag{44}$$

where $\chi$ is a learning rate hyperparameter. Equation (44) constitutes a simple first-order gradient descent method, and we note that more complex methods are available as is often employed for large parameter spaces.[66]

We conclude this theoretical section by investigating how the classical relative entropy emerges from its quantum definition. A direct but largely informative approach to bridge the quantum and classical definitions would be to take the expression for the relative entropy in Eq. (39), employ a discretization in terms of $P$ replicas as in Eqs. (25) and (27), and then to simply take the $P = 1$ limit. We will instead accomplish this quantum-classical bridging by posing the quantum relative entropy in terms of the most classical-like variable, the Feynman path centroid, which provides a deeper analysis.[67] For simplicity, we will assume distinguishability as we are strictly interested in the limiting case as classical behavior dominates. We will also operate in one spatial dimension; however the results presented can be readily generalized to multiple dimensions and particles. We first note that the entropy as expressed in Eq. (35) can be rewritten as an average over the path centroid in the following manner[27]

$$S_{\text{rel}} = - \langle \ln \rho_c \rangle_{\rho_c} \tag{45}$$

where for any operator $\hat{\mathcal{O}}$ we have defined the centroid average as

$$\langle \mathcal{O} \rangle_{\rho_c} = \frac{\int dr_c \rho_c(r_c)\, \mathcal{O}_c(r_c)}{\int dr_c\, \rho_c(r_c)} \tag{46}$$

In Eq. (46), we have incorporated the path centroid, $r_c = \left(\frac{1}{\hbar\beta}\right) \int_0^{\hbar\beta} d\tau\, r(\tau)$. We emphasize that in Eq. (46) the expression for the operator in the centroid representation, $\mathcal{O}_c$, is not necessarily the same as $\mathcal{O}$. This can be interpreted as a kind of representability problem for CGing the ring polymer to its centroid.[55] We note that the distributions considered here are for one degree of freedom and hence no dimensionality reduction in the CG model can be performed, i.e., the mapping entropy is zero. We thus refer to this 'CG' model merely as a separate thermal density matrix, $\hat{q}$. The relative entropy as defined in Eq. (39) can be expressed similarly using averages over the centroid ensemble

$$S_{\text{rel}} = \langle \ln \rho_c \rangle_{\rho_c} - \langle \ln q_c \rangle_{\rho_c} \qquad (47)$$

To proceed in the classical limit, we utilize an approximate expression for the centroid ensemble average of an operator[68]

$$\mathcal{O}_c(r_c) \cong \int d\tilde{r}\, \mathcal{O}(\tilde{r} + r_c) q(\tilde{r}) \qquad (48)$$

where $q(\tilde{r})$ is a Gaussian, defined as

$$q(\tilde{r}) = \frac{1}{\sqrt{2\pi \Delta r^2}} e^{-\tilde{r}^2/2\Delta r^2} \qquad (49)$$

which 'blurs' the classical centroid operator according to the quantum free particle thermal width $\Delta r^2 \approx \frac{\beta \hbar^2}{12m}$ (or a more accurate thermal width for a given potential). Note that from Eqs. (45) and (48) we have an approximate expression for the entropy of a quantum system

$$S \cong -\int dr_c\, \rho_c(r_c) \left[ \int d\tilde{r}\, \ln \rho_c(\tilde{r} + r_c)\, q(\tilde{r}) \right] \qquad (50)$$

We are interested in the behavior of this, albeit already approximate, expression for the entropy in the near classical regime. We begin by expressing $\rho_c(\tilde{r} + r_c)$ in terms of a Taylor series

$$S = -\int dr_c\, \rho_c(r_c) \left[ \int d\tilde{r}\, \ln \left( \rho_c(r_c) + \sum_{k=1}^{\infty} \frac{\tilde{r}^k}{k!} \partial_{r_c}^{(k)} \rho(r_c) \right) q(\tilde{r}) \right] \qquad (51)$$

We next manipulate the arguments of the natural logarithm in the following manner

$$S = -\int dr_c\, \rho_c(r_c) \left[ \int d\tilde{r}\, \ln \rho_c(r_c) + \ln \left( 1 + \frac{1}{\rho_c(r_c)} \sum_{k=1}^{\infty} \frac{\tilde{r}^k}{k!} \partial_{r_c}^{(k)} \rho(r_c) \right) q(\tilde{r}) \right] \qquad (52)$$

It is clear from Eq. (52) that Taylor expansion of the natural logarithm, followed by integration over $\tilde{r}$ will produce the moments of the distribution $q(\tilde{r})$. Our intent in this analysis is to examine the behavior of the quantum entropy as quantum effects diminish, i.e., where the moments of $q(\tilde{r})$ vanish. Consequently, let us assume the near-classical limit $\hbar \to 0$ and keep only leading order terms in the Taylor expansion of the natural logarithm

$$S = -\int dr_c\, \rho_c(r_c) \left[ \int d\tilde{r} \left( \ln \rho_c(r_c) + \frac{\tilde{r}}{\rho_c(r_c)} \partial_{r_c} \rho(r_c) + \frac{\tilde{r}^2}{2\rho_c(r_c)} \partial^2_{r_c} \rho(r_c) \right. \right.$$
$$\left. \left. - \frac{1}{2}\left( \frac{\tilde{r}}{\rho_c(r_c)} \partial_{r_c} \rho(r_c) \right)^2 \right) q(\tilde{r}) \right] + O(\hbar^4) \tag{53}$$

Following Eq. (53), we evaluate the first two terms in the Taylor series and keep only the leading order term. The first moment of $q(\tilde{r})$ is zero, hence we retain only the second order term

$$S = -\int dr_c\, \rho_c(r_c) \ln \rho_c(r_c) - \frac{\beta^2 \hbar^2}{24m} \int dr_c\, \rho_c(r_c) F_c'(r_c) + O(\hbar^4) \tag{54}$$

where in Eq. (54) we have defined the centroid force $F_c$ and its derivative $F_c' = dF_c/dr_c$. It is seen that the first term in Eq. (54) constitutes an entropy for the centroid distribution. We may write this as

$$S_c = -\int dr_c\, \rho_c(r_c) \ln \rho(r_c) \tag{55}$$

We can then simplify Eq. (54) to relate the quantum entropy to the centroid entropy plus a correction term

$$S = S_c - \frac{\beta^2 \hbar^2}{24m} \langle F_c'(r_c) \rangle_{\rho_c} + O(\hbar^4) \tag{56}$$

The results of Eq. (56) can readily be extended to the expression for the cross-entropy in Eq. (37). We assume that the model density matrix can be expressed as $\hat{q} \propto e^{-\beta \hat{h}_q}$. An expression for the semiclassical relative entropy can then be written as

$$S_{rel}[\hat{\rho}||\hat{q}] = S_{rel}[\rho_c||q_c] - \frac{\beta^2 \hbar^2}{24m} \langle F_c'(r_c) - F_{q,c}'(r_c) \rangle_{\rho_c} + O(\hbar^4) \qquad (57)$$

where $F_{q,c}$ denotes the derivative of the centroid force for the thermal density matrix $\hat{q}$. We emphasize that the first term on the right-hand side of Eq. (57) is the relative entropy between the centroid distribution functions. Additionally, it should be noted that the centroid distribution, as an effective distribution over all replicas, already has quantum effects and cannot be considered a purely classical object except in the classical limit. The centroid distribution reduces to the classical Boltzmann distribution in the classical limit, and thus the first term contains the purely classical analogue to the relative entropy.

The second term is a quantum correction term. It can be interpreted as an additional source of error between the FG and CG models which persists when the FG distribution changes appreciably on length scales of the thermal quantum width. Intriguingly, the leading order quantum correction term includes a kind of 'force-matching' term concerning the gradient of the centroid force. The semiclassical behavior of the quantum relative entropy, as shown in Eq. (57), is summarized in Fig. 2. It follows from the Taylor series in Eq. (51) that further terms in the series produce higher-order derivatives which would propagate into the error terms in Eq. (57). It is clear that the relative entropy is 0 when the FG and CG centroid distributions are identical, and that this condition holds for inclusion of any number of terms in the series of Eq. (51).

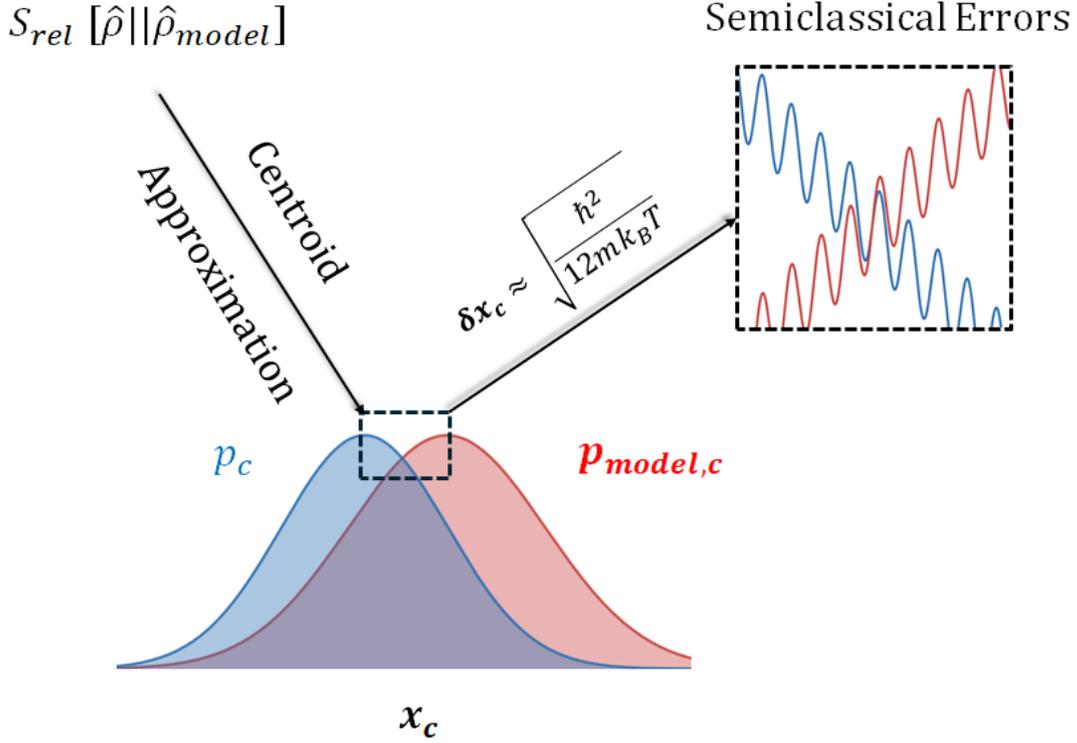

**Fig. 2.** Visualization of the semiclassical behavior of the relative entropy error metric. The quantum relative entropy (above, left) is first approximated via a centroid representation (middle, blue). A model centroid distribution is introduced (middle, red), and the subsequent error between the models includes the classical relative entropy errors as well as errors accounting for disparities at spatial scales of the thermal quantum width (right).

## III. APPLICATIONS

### A. Three-Dimensional Harmonically Trapped Bosons

**Path Integral Molecular Dynamics**. Bosonic PIMD simulations were conducted with i-PI 3.1.[69] We simulated $n = 3$ bosons with a mass of 1 electron mass unit which were confined in a three-dimensional harmonic trap of the following functional form

$$\hat{H}_{FG}(\hat{\mathbf{r}}_1, \hat{\mathbf{r}}_2, \hat{\mathbf{r}}_3, \hat{\mathbf{p}}_1, \hat{\mathbf{p}}_2, \hat{\mathbf{p}}_3 ; k) = \sum_{i=1}^{3} \frac{\hat{\mathbf{p}}_i^2}{2m} + \frac{1}{2} k(\hat{\mathbf{r}}_1^2 + \hat{\mathbf{r}}_2^2 + \hat{\mathbf{r}}_3^2) \tag{58}$$

The harmonic strength, $k$ was 0.027 kcal/mol Å$^2$. Note that for distinguishable particles, the resulting path integral Hamiltonian is separable into three one-particle Hamiltonians as particles do not interact with each other. However, for indistinguishable particles such as bosons, interparticle interactions are always present regardless of whether the potential explicitly includes such interactions due to exchange forces which introduce harmonic interaction terms between the ring polymers. The simulation temperature was 17.4 K and $P = 72$ replicas were used. A timestep of 1 fs was employed; time integration was conducted in Cartesian coordinates by necessity due to the lack of an analytical expression for the normal mode propagator of bosonic ring polymers. Due to the multiple timescales involved, multiple-time stepping was employed to separate potential and spring force calculations by a factor of 10 timesteps. The 3-boson system was simulated for 1.5 ns. A path-integral Langevin equation (PILE) thermostat was utilized to maintain the temperature with a friction constant of 100 fs.[70] To assess the degree of involvement of exchange effects throughout simulation, we calculate the probability for all bosons to be in distinct or 'unconnected' ring polymers

$$p_{dist} = \frac{1}{n!\, e^{-\beta V_B^n}} e^{-\beta \sum_{i=1}^n E_i^{(1)}} \tag{59}$$

where $E_i^{(1)}$ is the spring potential for all beads of particle $i$ to be connected as one ring. Across the simulation, the probability for all bosons to occupy a 'distinguishable' state was 17%, indicating very non-negligible exchange effects.

**REM.** We employed a center-of-mass (COM) CGing operation to collapse the FG system to 1 particle interacting in an effective harmonic well. The mass of this singular CG particle, $M$, was then chosen to be the sum of the masses of the FG system. The minimum of the well was held fixed at the origin and the harmonic strength was optimized via REM. The harmonic strength of the FG model was employed as an initial guess for the harmonic strength of the CG model.

$$\widehat{H}_{CG}(\hat{\mathbf{r}}_{COM}; K) = \frac{\hat{\mathbf{p}}_{COM}^2}{2M} + \frac{1}{2} K \hat{\mathbf{r}}_{COM}^2 \tag{60}$$

The CG system was simulated at the same temperature and number of replicas as the FG system. Each REM iteration was simulated for 50 ps at a timestep of 1 fs utilizing a PILE thermostat with a friction constant of 100 fs to maintain the system temperature. Gradient descent, as shown in Eq. (44), was utilized with a learning rate of 0.0002 for 34 iterations.

**Results.** We utilized the gradient of the relative entropy, shown in Eq. (42), to assess convergence during training. Fig. 3 demonstrates the behavior of the relative entropy gradient throughout training, and is indicative of convergence to a minimum. The final spring constant obtained was $K = 0.011$ kcal/mol Å$^2$.

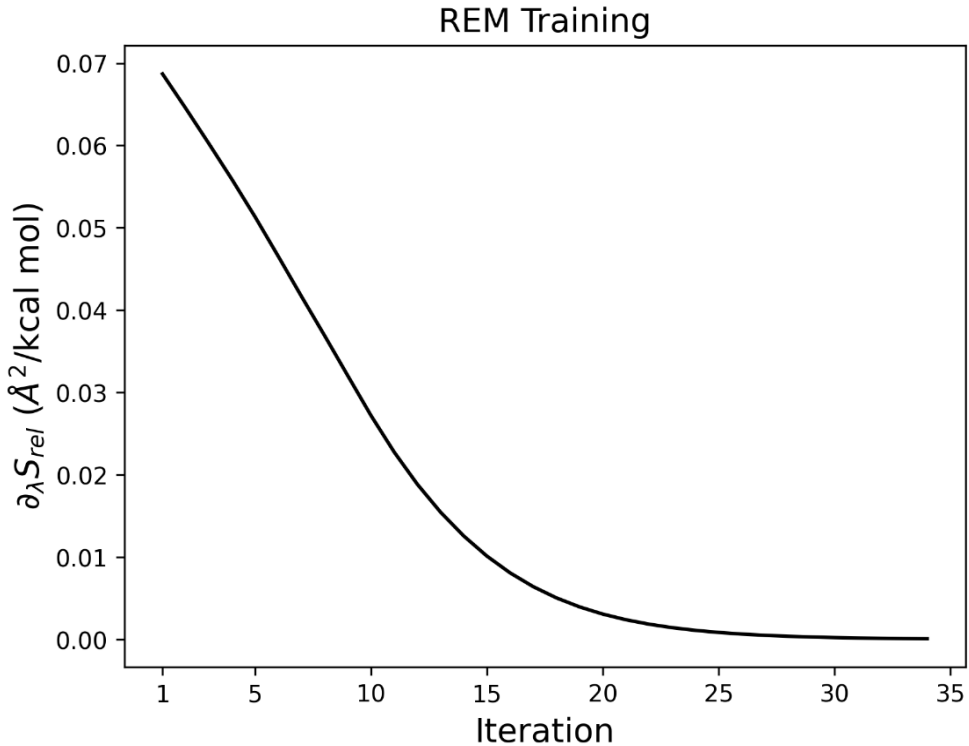

**Fig. 3.** Numerical convergence in REM training of CG model of harmonically trapped bosonic system.

The spatial distributions of the CG system are shown in Fig. 4a) and Fig. 4b). As expected via Eq. (42), the average square distance of the COM between the FG and CG models are equivalent as shown in Fig. 4a). However, as has been observed previously in classical implementations of the REM method,[71] this recapitulation of average behavior does not necessarily come with capturing higher-order moments in the distribution. It is clear from Fig.

4a) and Fig. 4b) that the CG model tends towards greater delocalization than the FG model. This can be ameliorated in two ways. First, in principle, a completely expressive basis set should recapitulate all statistics of the exact CG model. However, on a deeper level, the results of Fig. 4) suggest, commensurate with previous results,[26] that the assumption that the CG Hamiltonian is local in imaginary time is not adequate. Instead, non-local influence functionals such as those utilized in the study of open quantum systems[72] could be employed as a quantum CG "basis set" to better capture delocalization effects in CG systems.

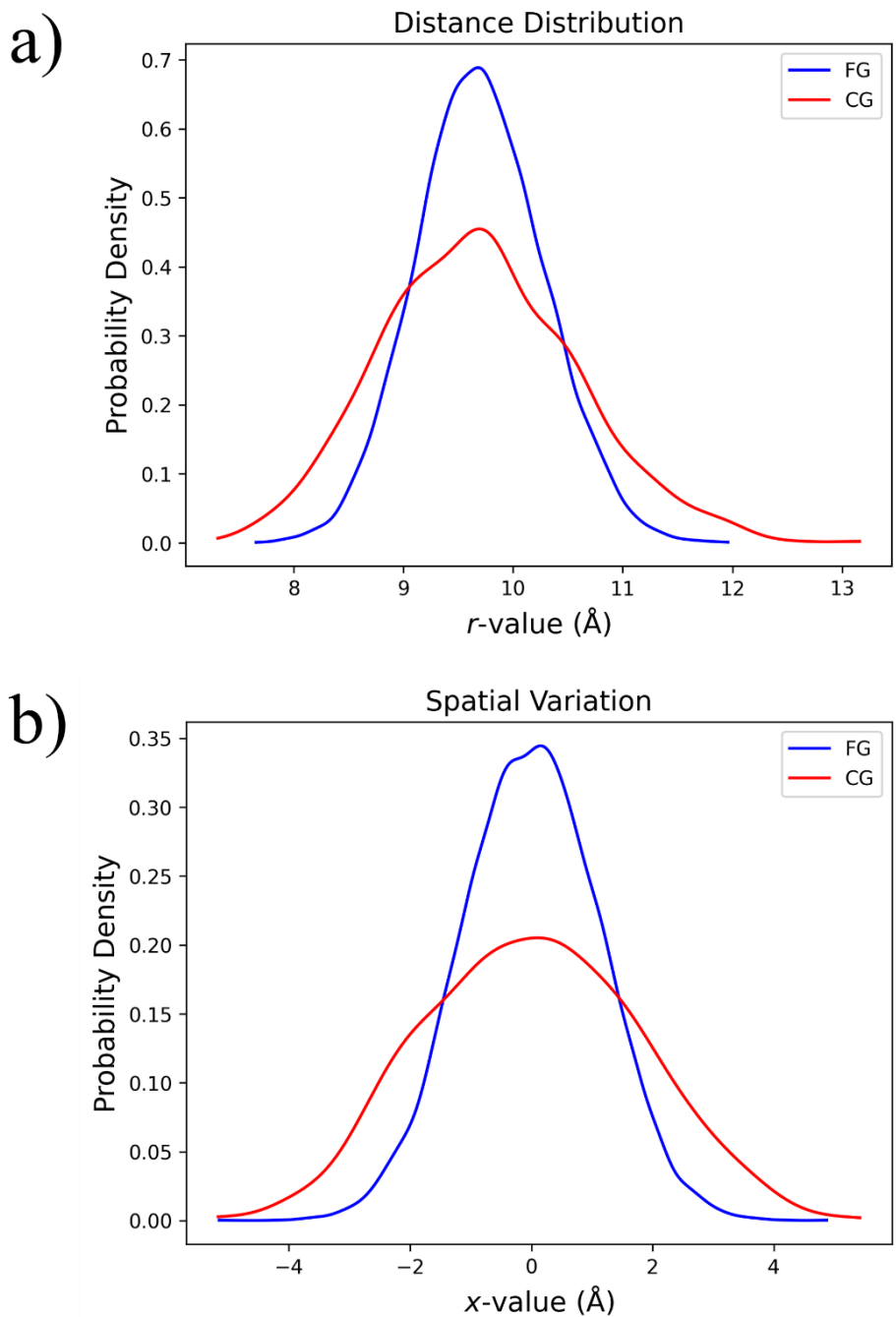

**Fig. 4.** Distribution of a) distance and b) spatial x-value for the center of mass of the bosonic system for the FG and CG models.

**B. Quantum Coarse-Graining of Liquid Water**

**Path Integral Molecular Dynamics.** All simulations were conducted utilizing the i-PI software package with LAMMPS for force evaluation.[73, 74] The atomistic system consisted of 233 water molecules simulated using the qSPC/Fw force field[75] at a temperature of 300 K and $P = 32$ replicas in a cubic box with a length of 19.034 Å. The atomistic system was simulated for 225 ps at a timestep of 0.5 fs. A PILE thermostat with a friction constant of 25 fs was utilized to maintain the temperature.

**REM.** A COM CGing operation was used for mapping of each water molecule. The OpenMSCG software package was utilized for mapping, modeling, and analysis.[76] The one-site CG water pairwise interaction potential was parameterized via the following functional form

$$\hat{U}_{CG}(\hat{\mathbf{R}}; \boldsymbol{\theta}) = \sum_{J<I} \hat{U}_{IJ}(\hat{\mathbf{R}}_I, \hat{\mathbf{R}}_J) = \sum_{J<I} \sum_{k} \theta_k B_k^l (|\hat{\mathbf{R}}_I - \hat{\mathbf{R}}_J|) \tag{61}$$

where $B_k^l$ is a $l^{th}$ order B-spline and the sum over $k$ constitutes a sum over all B-spline components. In this work, we employ $4^{th}$ order B-splines with a resolution of 0.3 Å to fully define the interaction potential of Eq. (61). We then conducted REM optimization over the set of B-spline coefficients $\boldsymbol{\theta}$. An initial guess potential was obtained via direct Boltzmann inversion (BI) of the radial distribution function (RDF).[77] The RMSProp algorithm was then employed for 80 iterations of gradient descent.[78] A learning rate of 0.08 and a decay rate of 0.99 was employed for 80 iterations. The first and last knots were held fixed during optimization to maintain a hard-wall interaction at low distances and a decay near the cutoff, respectively. The mass of the CG water bead was set to the sum of the mass of all atoms in a water molecule. Each REM iteration consisted of 25,000 MD timesteps with a timestep of 0.5 fs, and a PILE thermostat with a friction constant of 25 fs was used to maintain the system temperature.

**Results.** The BI potential and the final REM potential are shown in Fig. 5. Notable features that REM optimization induces are an attractive well from the 4-5 Å region and a repulsive region from 3-4 Å. An attractive well at roughly 2.8 Å is maintained for both models.

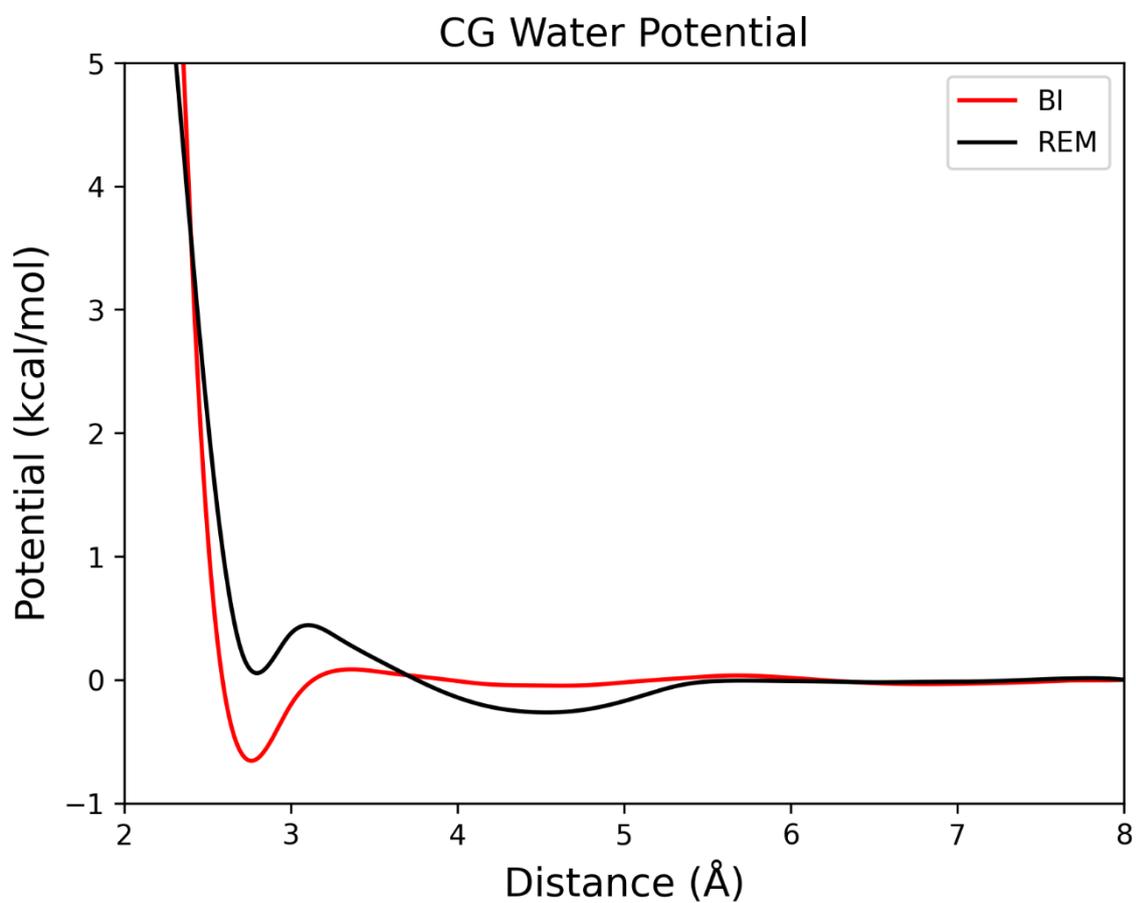

**Fig. 5.** Potentials of the BI and REM CG water models.

We plot the two-body correlation at the CG resolution for the mapped all-atom (AA), BI, and REM CG models in Fig. 6. It is clear that the REM model improves upon the initial BI potential. As expected from behavior observed in classical REM optimization,[79] the spline potential is flexible enough to capture quantitative aspects of the RDF including the first solvation peak.

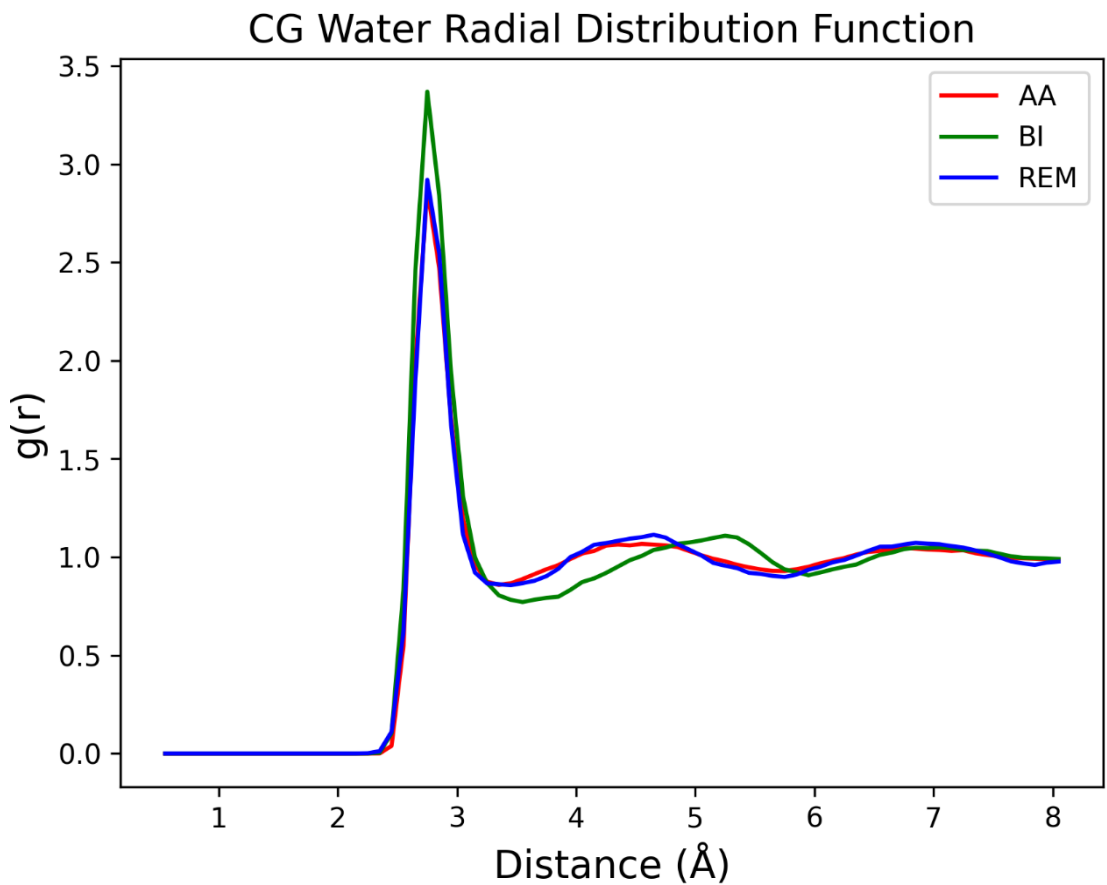

**Fig. 6.** Radial distribution functions of the mapped AA, BI, and REM CG water models.

We next plot the three-body correlations at the cutoff threshold of 3.3 Å in Fig. 7. It is seen from Fig. 7 that pairwise modeling of one-site CG water cannot capture higher-order correlations, which is behavior also observed in CG modeling of classical water.[80] We emphasize that this is not a shortcoming of the REM optimization method but rather of the descriptive capabilities of the CG force-field basis set. Consequently, higher-order terms such as Stillinger-Weber three-body interactions or more expressive machine-learning potentials are ultimately necessary.[81]

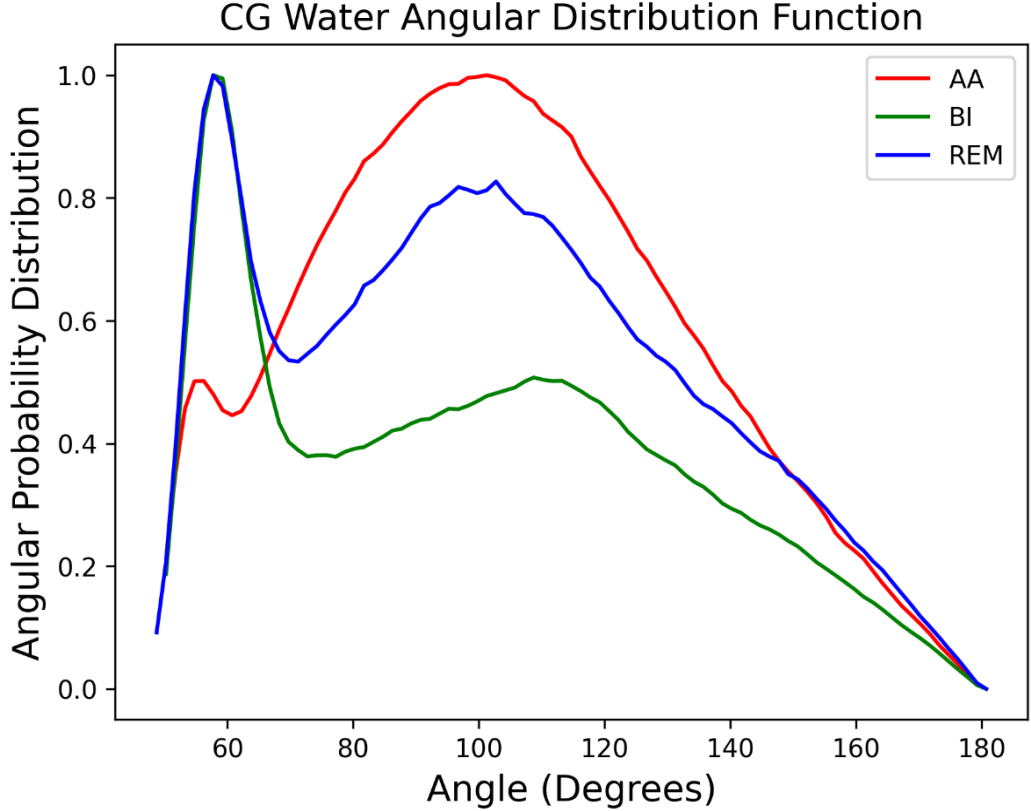

**Fig. 7.** Angular distribution functions of the mapped AA, BI, and REM CG water models.

## IV. DISCUSSION

In this work, we have investigated the particle-based CGing of an indistinguishable bosonic system via a newly developed quantum REM principle, from which numerous alternative CGing strategies may emerge. Future directions may explore 'CGing' exchange effects rather than the number of interacting particles to reduce the computational burden which arises from particle indistinguishability. Such 'CG' models could be trained via the REM algorithm described by Eq. (44) through conventional PIMD. This approximate CGing method may be fruitful for systems with only weakly quantum degenerate behavior. An alternative route towards CGing, as discussed in the Introduction, is to CG the intermediate imaginary time slices or beads. This would allow for reductions in particle number while still incorporating exchange effects explicitly in the Hamiltonian, as particle exchange only occurs in the 'final' imaginary time slice.

Given advances in condensed phase simulation of bosonic systems, a promising application of bosonic CG modeling is to capture the behavior of superfluids such as He$^4$ at low temperatures. We expect such CG modeling to be particularly fruitful given the previous successes of bottom-up CG methods in recapitulating condensed phase molecular behavior[82-84] and for CG modeling to significantly push forward the spatial scales of bosonic PIMD. CG modeling of interacting fermions, however, remains complicated by the sign problem of many-body fermionic systems, which precludes large-scale PIMD simulation of fermions. We note, however, that so long as the expectation value in Eq. (42) can be evaluated, REM can be conducted for 'CGing' of fermionic systems. For a small number of weakly quantum degenerate fermions, this can be accomplished via bosonic PIMD simulation and subsequent 'reweighting' to account for sign changes in the fermionic PIMD Hamiltonian.[61] Alternatively, an analytic continuation of $\xi$ can be performed such that information in the fermionic regime, i.e., $\xi < 0$, can be extrapolated via an approximate phenomenological equation for the variation in observable expectation values, $\langle \hat{O} \rangle(\xi)$.[85] However, it has been observed that such approximations break down with increasing quantum degeneracy.[86] Consequently, it is expected that, at the present time of this work, CGing of fermions can feasibly be accomplished only for weakly quantum degenerate systems, whereas CGing of moderate to strongly quantum degenerate fermionic systems will ultimately necessitate further advancements in combating the sign problem.

While we have presented a distribution-matching based approach towards CGing of indistinguishable particles in this work, force-based methods such as the MS-CG method have not been formally developed here or explored for indistinguishable particles. A particular advantage of force-based methods over distribution-based methods such as REM is their non-iterative nature; force-based methods carry out gradient descent in parameter space without necessitating repeated CG simulation.[87] This is particularly advantageous for implementation of machine-learning potentials, which are often more computationally demanding than 'classical' force-field models. Due to the necessity of incorporating many-body terms in capturing complete configurational statistics, as demonstrated in this work, we expect the utilization of ML potentials here to be valuable. However, we do note that force-matching based methods typically restrict the mapping operation to linear mapping operators,[88] limiting the form of CG mapping permitted.

The semiclassical expression for the relative entropy as detailed in Eq. (57) provides an intriguing alternative to the fully quantum expression employed in this work. The quantum REM algorithm herein necessitates PIMD simulation of both the FG and CG model, where simulation of the latter is conducted iteratively for gradient descent. Instead, the semiclassical expression of Eq. (57) concerns a variationally optimal definition of the CG centroid distribution which can be sampled from conventional MD simulation without bead replicas. Consequently, the form of Eq. (57) implies a gradient descent procedure where one conducts PIMD of the FG model to extract FG centroid statistics and iteratively simulates the CG centroid model via classical MD to conduct gradient descent. We reserve the theoretical expansion of Eq. (57) to include CG mappings and its numerical analysis for future work.

## V. CONCLUDING REMARKS

In this work, we have expanded upon the formal theoretical methodology of bottom-up CGing to incorporate quantum mechanical indistinguishable particles. We have additionally established the quantum analogue of the relative entropy training algorithm for variational optimization of bottom-up CG models as well as its semiclassical approximation in terms of the Feynman path centroid representation. We have demonstrated that, similar to its classical counterpart, the quantum REM method reproduces FG correlations in the CG model which are dual to the interaction terms optimized. Future directions of this work include CG modeling of condensed phase bosonic systems, CG modeling of weakly degenerate fermionic systems, and the application of ML models in CG modeling of such systems due to their many-body interaction character.

ACKNOWLEDGEMENTS

This material is based upon work supported by the National Science Foundation (NSF grant CHE-2102677). Simulations were performed using computing resources provided by the University of Chicago Research Computing Center (RCC).

DATA AVAILABILITY

Data available on request from the authors.